\documentclass[twocolumn,superscriptaddress,pra,aps]{revtex4}
\usepackage{graphicx}
\usepackage{amsmath}
\usepackage{amssymb}
\usepackage{epstopdf}
\usepackage{color}

\begin{document}
\title{Geometric extension of Clauser-Horne inequality to more qubits}
\author{Arijit Dutta}
\affiliation{School of Computational Sciences, Korea Institute for Advanced Study, Seoul 02455, Korea}
\affiliation{Department of Physics, Hanyang University, Seoul 04763, Korea}
\author{Tschang-Uh Nahm}
\author{Jinhyoung Lee}
\email{hyoung@hanyang.ac.kr}
\affiliation{Department of Physics, Hanyang University, Seoul 04763, Korea}
\author{Marek \.Zukowski}
\affiliation{Institute of Theoretical Physics and Astrophysics, University of Gda\'{n}sk, 80-952 Gda\'{n}sk, Poland\\ and 
Faculty of Physics, University of Vienna, A-1090 Vienna, Austria}

\newcommand{\ket}[1]{|#1\rangle}
\newcommand{\bra}[1]{\langle #1|}
\newcommand{\ketbra}[2]{|#1\rangle \langle #2|}
\newcommand{\braket}[2]{\langle #1|#2\rangle }
\newcommand{\avg}[1]{\langle #1\rangle}

\begin{abstract}
 We propose a geometric multiparty extension of  Clauser-Horne (CH) inequality. The standard CH inequality can be shown to be an implication of the fact that statistical separation between two events, $A$ and $B$, defined as $P(A\oplus B)$, where $A\oplus B=(A-B)\cup(B-A)$, satisfies the axioms of a distance.  Our  extension for tripartite case is based on triangle inequalities for the statistical separations of three probabilistic events $P(A\oplus B \oplus C)$. We show that Mermin inequality can be retrieved from our extended CH inequality for three subsystems. With our tripartite CH inequality, we investigate quantum violations by GHZ-type and W-type states. 
Our inequalities are compared to another type, so-called $N$-site CH inequality. In addition we argue how to generalize our method for more subsystems and measurement settings. Our method can be used to write down several Bell-type inequalities in a systematic manner.

\end{abstract}

\maketitle

\section{Introduction}
Intrinsic randomness of quantum mechanics has been a  topic of a debate  for many years. In 1935, Einstein, Podolsky, and Rosen (EPR) claimed  that quantum mechanics is an incomplete theory \cite{EPR35}, and hidden variables could be implemented to resolve the issue. In his pioneering work~\cite{Bell64} Bell formulated an inequality that is satisfied by local hidden variable models, but can be violated by quantum mechanics for bipartite and two-level systems. 
 Bell-theorem significantly improved our understanding of quantum intrinsic randomness with respect to the assumptions of reality and locality in local hidden variable models. Further studies have been done by considering more complicated systems. For examples, there have been Bell inequalities proposed for many qubits with two dichotomic observables per site~\cite{Mermin1990, Werner2001, weinfurter2001, Zukowski2002, guney2013} and for more than two alternative observables~\cite{wu2003, laskowski2004}. For higher dimensional systems, different forms of Bell inequalities have been introduced~\cite{mermin1980, mermin1982, ardehali1991, garg1982, wodkiewicz1994, chen2001, Kaszlikowski00, kaszlikowski2002, CGLMP02, zohren2008, gruca2012, Son06a}. Other derivations of Bell inequalities can be found also in Ref.~\cite{Brunner2014}. 

 Experimental tests to invalidate Bell inequalities  faced challenges such as detection and locality loopholes ~\cite{Brunner2014}. In particular, the detection loophole had to be dealt with by the use  with the additional fair-sampling assumption, i.e., that the detected events give a fair representation of the entire ensemble \cite{pearle1970,Clauser74}.  In 1974, Clauser and Horne proposed another type of Bell inequality~\cite{Clauser74}, which is very handy in dealing with such problems like detection inefficiency (see e.g the analysis by Eberhard ~\cite{eberhard1993, khrennikov2014a}), and can be thought of as the most elementary Bell inequality. Some attempts to generalize the inequality to multipartite systems can be found in Refs.~\cite{larsson2001, cabello2002b}. We shall continue here this effort, however our generalization will be based on different observations concerning the original CH inequality and therefore will take a different form. Our extension is based on a geometric interpretation of Bell inequalities in Kolmogorov theory of probability  as given by ~\cite{santos1986}, further developed in e.g. Refs. \cite{pykacz1991, zukowski2014}.
 
 An experimental falsification of Bell's inequality without the fair-sampling assumption was recently presented in Ref.~\cite{giustina2013}. Recently, a loophole-free violation of Bell inequality was reported by Hensen et al.~\cite{hensen2015}, even though many attempts were previously made to close the two loopholes simultaneously (see Ref.~\cite{pan2012} for important results). The possibility of having loophole free realizations of Bell experiments, opens the way for constructing device independent quantum information schemes and protocols, which would be fully secure. However,  thus far we do not have a loophole free Bell experiment for three or more subsystems, and therefore one must continue research toward  finding optimal approaches in this realm.

 The essential traits of our generalization are as follows:
 
 \begin{itemize}
 	\item 
We first note that at the times in which CHSH and CH inequalities were first formulated, they looked as ad-hoc ones. Their starting points were certain, indeed ad-hoc, algebraic identities. Still as 
  the CHSH-Bell inequality is implied by the CH one, the latter one seemed from the very beginning to be more fundamental. Much later, see ref. \cite{santos1986}, it turned out that the  CH inequality is derivable using a geometric notion of separation of two probabilistic events. The separation for two events $A$ and $B$ reads $S(A,B)=P(A)+P(B)-2P(A,B)$.  It has all properties of a distance, this includes most importantly the triangle inequality. The triangle inequality cannot be used to derive Bell inequalities, still suitable chaining of two triangle inequalities 
  leads to a quadrangle one which, after it is re-written in terms of probabilities is the Clauser-Horne inequality. This geometric feature, we think, underlines the fundamental nature 
  of the CH inequality, and singles it out. Therefore we postulate that three (or more) Bell inequalities which are direct generalizations of
 the CH one should be derivable also using the geometric properties of separations of three (or more) probabilistic events.
\item 
 The  statistical separation between probabilistic events, $A$ and $B$, can also be put as $P(A\oplus B)$, where $A\oplus B=(A-B)\cup(B-A)$. We find an extension of CH inequality for a tripartite system, by an extension of the  separation measure to $P(A\oplus B \oplus C).$ Most importantly we have $A\oplus (B \oplus C)=(A\oplus B) \oplus C=A\oplus (C \oplus B)$.
 The symmetry property of the two event separation, $S(A,B)=S(B,A)$, plays no obvious role in the derivation of the CH inequality.
 However its extension, shown above in the three party scenario, is essential to derive the initial triangle inequalities, with the use of which one can derive our generalization of  the CH inequality for a tripartite system.
\item 
 The inequality can used to derive the  Mermin inequality for three qubits~\cite{Mermin1990}, and of course, further on to the  CHSH inequality of two qubits in Ref.~\cite{Clauser74,CHSH69}. So the generalization  has an additional trait similar to the CH inequality. We show also that it can be reduced to a bipartite CH inequality if one party is eliminated. 
 \item The inequality is tight, just as the CH one.
 \item
 The method can be further extended to  more than three parties, however in this case non-trivial generalizations (along the lines: a polygon inequality for separations)
of the CH inequality, in the case in which the parties play fully symmetric roles,  involve more than two settings for each party.
\end{itemize}

The paper is organized as follows. We derive a geometric extension of CH inequality for three subsystems by introducing statistical separation of probabilistic events in Sec.~\ref{section1}. There we also discuss the properties of the inequality such as it being tight, and reducible to a bipartite CH inequality and a tripartite Mermin inequality. In Sec.~\ref{section3}, quantum violations of our  inequality and $N$-site CH inequality f Ref.~\cite{larsson2001} are discussed in terms of the degree of noise robustness and critical efficiency of detection. There is region of parameters in which our approach gives quantum violations which are more resistant to the imperfections. We also demonstrate the generality  of our method by extending the inequality for a three-party system with two measurement settings per cite to a three-party system with three measurement settings per site and a four-party system with three measurement settings per site (Sec.~\ref{ext}). We summarize our results in Sec.~\ref{section5}.

\section{derivation}
\label{section1}
\subsection{Geometric approach with statistical separations}
\label{sec:subII1}
In our derivation of  CH-type Bell inequalities, statistical separation plays a crucial role. It is defined by the probability of symmetric difference between two events ~\cite{Renyi70}. Let $(\Omega, F, P)$ be a probability space, where $\Omega$ is a sample space, $F$ an event space, and $P$ a probability measure. The symmetric difference of two  events $X$ and $Y$ is defined by
$$ X \oplus Y = (X-Y)\cup (Y-X) = X \cap Y^c \cup X^c \cap Y,$$
where $X^c=\Omega -X$, etc.
It satisfies the following properties: (a)  $X \oplus X = \emptyset,$ (b)   $X \oplus Y = Y \oplus X$, and (c) it is associative, $(X\oplus Y) \oplus Z = X \oplus (Y \oplus Z)$.
For three events we have, most importantly, the permutation symmetry,   that is  $X\oplus (Y \oplus Z) = Y\oplus (Z \oplus X) = Z\oplus (X \oplus Y)$. 
The generalization to more events is inductive. As events  $X \cap Y^c$ and  $X^c \cap Y$ are mutually exclusive  one has $1\geq P(X\oplus Y)=P(X \cap Y^c) + P(X^c \cap Y)\geq 0.$ The probability
$P(X \oplus Y)$ of a symmetric difference is often called statistical separation between  events $X$ and $Y.$ It was originally put as 
\begin{equation}
\label{eq:dss}
P(X \oplus Y) = P(X) + P(Y) - 2 P(X, Y),
\end{equation}
where $P(X, Y)\equiv P(X\cap Y).$
 
 Statistical separation has a geometric interpretation as it obeys the triangle inequality,
\begin{equation}
\label{ttt}
 P(X \oplus Y)+P(Y \oplus Z)\geq P(X \oplus Z). 
\end{equation}
Note that  for events $X, Y,$ and $Z$ one has $(X \oplus Y) \oplus (Y \oplus Z) = X \oplus Z.$ The inequality (\ref{ttt}) in terms of probabilities reads
\begin{equation}
\label{ttt1}
 P(Y)+P(X, Z)\geq P(X, Y)+P(Y, Z).
\end{equation}

Let us consider probability $P(X \oplus Y \oplus Z)$ of the measure of the symmetric difference of three events, $X, Y,$ and $Z.$ It because of the aforementioned permutation symmetry is a statistical separation between $X$ and $Y \oplus Z,$ or between $Y$ and $Z \oplus X,$ or  finally between  $Z$ and $X \oplus Y.$ We define it as  a statistical separation of three events $X, Y,$ and $Z$. One can show that $X \oplus Y\oplus Z \equiv (X^c \cap Y^c\cap Z) \cup (X \cap Y^c\cap Z^c) \cup (X^c \cap Y \cap Z^c) \cup (X \cap Y\cap Z).$ By noting $X^c \cap Y^c \cap Z,$ $X \cap Y^c \cap Z^c,$ $X^c \cap Y \cap Z^c,$  and $X \cap Y \cap Z $ are mutually exclusive, we also get
\begin{eqnarray}
\label{KOL1}
P(X\oplus Y \oplus Z)&=& P(X, Y, Z)+P(X, Y^c, Z^c)\nonumber \\
&+& P(X^c, Y^c, Z)+P(X^c, Y, Z^c),
\end{eqnarray}
where $P(X, Y, Z) \equiv  P(X \cap Y \cap Z),$ and so on. Such considerations can be extended to more events that three.

In the following subsection,  we analyze statistical separations of certain combinations of symmetric differences of probabilistic events for a specific type of three-qubit experiments.

\subsection{A geometric tripartite extension of CH inequality}
\label{chmm}
  Here we derive  a Bell-type inequality for  statistical separations for tripartite systems. We consider a scenario of three qubits. The qubits are measured by three observers (Alice, Bob, and Charlie). Each partner choose between two measurement settings. Events associated with Alice's (Bob's and Charlie's) choice of measurement settings $i$ are denoted by $A_i$ ($B_i$ and  $C_i$), for $i=1,2$. For example, event $A_i$ for Alice represents the fact that Alice chooses a measurement related with  a {\em projector} $\hat{A}_i$ and detects her qubit.  We define the events $B_j$ and $C_k$ for  Bob and Charlie, respectively, in a similar way. The respective outcomes, denoted as  $a_i$ $b_j$ and $c_k,$,  mean `detection' if their  value is 1 and `no detection' is 0, i.e. $a_i, b_j, c_k= 0,1$. 
 
 In local realistic probabilistic model, the outcomes build  elements of the  sample space for the considered scenario $\Omega = \{(a_1,a_2,b_1,b_2,c_1,c_2) \,|\, a_i, b_j, c_k = 0,1\}$. Note that is such a treatment, we avoid introduction of other hidden variables. Only the possible full sets of hidden results suffice.
 Events $E$ are subsets of the sample space, $E \subseteq \Omega$, and their probabilities are denoted by $P(E).$ For instance, $P(A_1)$ is the probability of event 
 $$A_1 = \{(a_1=1,a_2,b_1,b_2,c_1,c_2) \,|\, a_2, b_{1,2}, c_{1,2} = 0,1 \},$$ 
that is, the detection of Alice's qubit with the choice of projector $\hat{A}_1.$   
Joint probabilities of two events $A_i$ and $B_j$, that Alice's and Bob's qubits are all detected for local settings $i$ and $j$, are denoted by $P(A_i, B_j).$ This applies to the other pairs, Bob and Charlie, and Charlie and Alice. Similarly, detections of all qubits by the observers (Alice, Bob, and Charlie) in settings $(i, j,$ and $k,$ respectively) are associated with joint probability  $P(A_i, B_j, C_k)$ of the three detection events $A_i$ and $B_j$ and $C_k.$ 

Note that the following triangle inequality holds 
\begin{eqnarray}
P\big((A_1\oplus B_2)\oplus C_2\big) + P\big(C_2\oplus (A_2\oplus B_1\big) \big)\nonumber \\ \geq P\big((A_1\oplus B_2)\oplus (A_2\oplus B_1)\big).
\end{eqnarray}
In the above the additional inside brackets we introduced to point that we in fact deal here with the original triangle inequality for separation of pairs of events (\ref{ttt}).
However they can be dropped, as we have the symmetries mentioned earlier. The second triangle inequality which we need is 
\begin{eqnarray}
P\big((A_1\oplus B_1)\oplus(A_2\oplus B_2)\big) + P\big((A_2\oplus B_2)\oplus C_1\big) \nonumber \\ \geq P\big((A_1\oplus B_1)\oplus C_1\big),
\end{eqnarray}
 and it is  written in a similar convention. Note that both inequalities involve events which form the quantum point of view  are inaccessible experimentally, if the projectors associated with $A_1$ and $A_2$  (also $B_1$ and $B_2$) do not commute. 

With the two triangle inequalities we get an inequality which applies only to operationally accessible situations
\begin{eqnarray}
\label{tttt12}
&& P(A_1\oplus B_2\oplus C_2) + P(A_2\oplus B_1\oplus C_2) \nonumber \\
&& \quad + P(A_2\oplus B_2\oplus C_1) \ge P(A_1\oplus B_1\oplus C_1).
\end{eqnarray}
This is the geometric tripartite extension of CH inequality.

To get a form of it which is expressed in terms of probabilities of events and their coincidences, we expand this inequality by using Eq.~\eqref{eq:dss}, to get 
\begin{eqnarray}
\label{CHM}
L + Q_{AB} + Q_{BC} + Q_{CA} + 2T \geq 0,
\end{eqnarray}
where $L$ is the sum of local (single-site) probabilities, $Q_{XY}$ and $T$ are  certain combinations of pair- and triple-site joint probabilities, respectively. More explicitly,
\begin{eqnarray*}
L&=& \sum_{X=A,B,C} P(X_2) \nonumber\\
Q_{XY} &=& P(X_1,Y_1) - P(X_1, Y_2) -P(X_2,Y_1) - P(X_2,Y_2) \nonumber \\
T&=& P(A_1, B_2, C_2)+ P(A_2, B_1, C_2)+ P(A_2, B_2, C_1) \nonumber \\
&& -P(A_1, B_1, C_1).
\end{eqnarray*}


The tripartite inequality~$\eqref{CHM}$ is reduced to a bipartite CH inequality if one party is eliminated. Assume that Charlie has no events of detections in his measurements, i.e. all probabilities of events $C_{1,2}$ vanish. Then, $L = P(A_2) + P(B_2)$, $Q_{AB} = P(A_1,B_1) - P(A_1, B_2) - P(A_2,B_1) - P(A_2,B_2)$, $Q_{BC} = Q_{CA} = 0$, and $T = 0$. In other words, the tripartite inequality in Eq.~\eqref{CHM} becomes a bipartite CH inequality, if the third party does not measure anything, i.e. its events are represented in the sample space by empty sets ~\cite{Clauser74}. It is worth to mention that the left hand side in Eq.~\eqref{CHM} is upper bounded by $1.$

We may formulate the inequality~$\eqref{CHM}$ in a form of Eberhard inequality for 3 qubits, as shown in Appendix~\ref{sec:eche1}. Eberhard inequality~\cite{eberhard1993} includes explicitly non-detection events. Its derivation from CH inequality is presented in Appendix~\ref{sec:eche}, showing the algebraic equivalence between the two inequalities~\cite{khrennikov2014a}.

\subsection{Mermin inequality}
\label{rm}
Our method can be applied to deriving Mermin inequality, just like CH ones lead to CHSH ones. Let us consider an experiment where each of three observers possesses a two-channel analyzer and two detectors, each with two outcomes $(\pm 1)$. Suppose the experiment is described by a local hidden variable model with sample space $\Omega = \{(a_1,a_2,b_1,b_2,c_1,c_2) \,|\, a_i, b_j,c_k =\pm 1\}.$ In this scenario we assume that every subsystem is detected with one of two outcomes $\pm 1$, contrary to the previous subsections. Event $A_1$ is defined by the detection with outcome $+1$ of Alice's qubit for a projector $\hat{A}_1$ chosen and event $A_1^c$ by the one with outcome $-1.$ That is, events
$$A_1 = \{(a_1=+1,a_2,b_1,b_2,c_1,c_2)|a_2,b_{1,2},c_{1,2}=\pm 1\}$$ and $$A_1^c = \{(a_1=-1,a_2,b_1,b_2,c_1,c_2)|a_2,b_{1,2},c_{1,2}=\pm 1\}.$$ Similarly, other events are defined. To this end, we obtain an inequality, as in Eq.~\eqref{tttt12}, by applying the procedure similar to Sec.~\ref{chmm}. The inequality, in the form of Eq.~$\eqref{tttt12}$, is then Mermin inequality. To show this, we construct correlation functions in terms of probabilities. Note first that $(X \oplus Y)^c = X \oplus Y^c = X^c \oplus Y$ so that $X \oplus Y \cup X \oplus Y^c = \Omega$. This implies $P(A \oplus B \oplus C) + P(A \oplus B \oplus C^c) = 1,$ i.e. a normalization condition. The correlation functions are given by 
$$
E_{\alpha_i \beta_j \gamma_k}= P(A_i\oplus B_j\oplus C_k)-P(A_i\oplus B_j\oplus C^c_k),
$$ 
where $\alpha_i, \beta_j,$ and $\gamma_k$ parameterize the local settings, respectively. Note from Eq.~\eqref{KOL1} that $P(A\oplus B\oplus C)$ contains the joint probabilities of even numbers of outcomes with $-1$, while $P(A\oplus B\oplus C^c)$ does the ones of odd numbers of outcomes with $-1$, so that  $E_{\alpha_i \beta_j \gamma_k}$ are the same correlation functions as those in Ref.~\cite{Mermin1990}. Applying the normalization condition, we obtain  $E_{\alpha_i \beta_j \gamma_k}=2P(A_i\oplus B_j\oplus C_k)-1.$ Replacing $P(A_i\oplus B_j\oplus C_k)$ with $\left(1+E_{\alpha_i \beta_j \gamma_k}\right)/2$ in Eq.~$\eqref{tttt12}$, we obtain Mermin inequality~\cite{Mermin1990},
\begin{eqnarray*}
E_{\alpha_1 \beta_2 \gamma_2} + E_{\alpha_2 \beta_1 \gamma_2} + E_{\alpha_2 \beta_2 \gamma_1} - E_{\alpha_1 \beta_1 \gamma_1} \ge -2.
\end{eqnarray*}
Here, we have the lower bound of Mermin inequalities. The upper bound $2,$ can be obtained in a similar way.

\subsection{$N$-site CH inequality}
\label{section2}
In this subsection we briefly discuss one of the multipartite extensions of CH-inequality, so-called a $N$-site CH inequality that Larsson and Semitecolos proposed~\cite{larsson2001}. For three qubits with $N=3$, the 3-site CH inequality reads
\begin{eqnarray}
\label{CHMM}
 T_1-T_2-Q_1 \leq 0,
\end{eqnarray}
where $Q_1$ is the sum of a specific set of pair-site probabilities, and $T_1$ and $T_2$ are of triple-site probabilities. More explicitly,
\begin{eqnarray*}
Q_1 &=& P(A_1,B_1) + P(B_1, C_1) +P(A_1, C_1) \nonumber \\
T_1&=& P(A_1, B_1, C_2)+ P(A_1, B_2, C_1)+ P(A_2, B_1, C_1)\nonumber\\&&+2P(A_1, B_1, C_1)\nonumber \\
T_2&=& P(A_1, B_2, C_2)+ P(A_2, B_1, C_2)+ P(A_2, B_2, C_1).\nonumber
\end{eqnarray*}
They proposed the $N$-site CH inequality to find the minimum detection efficiency required for quantum mechanics to violate the inequality. 
In later sections, we will discuss the results related to the $3$-site CH inequality and compare them to our geometric tripartite extension of CH inequality.

\subsection{Tightness of geometric tripartite extension of CH inequality}
We show that our geometric tripartite extension of CH inequality is tight, i.e. it is a facet inequality of a local realistic polytope so that it can discriminate sharply the domains of the local realistic and quantum correlations~\cite{Brunner2014}. For the purpose we consider the local realistic polytope with the experimental scenario in which the inequality is derived. The local realistic polytope, called a Bell polytope $\mathcal{B}$, is a collection of probability vectors
$$
\vec{P}=\sum_{i} \lambda_i \vec{p}_i,
$$ 
where $\vec{p}_i$ are vertices of the polytope (or extremal points), $\lambda_i$ are positive real numbers and  $\sum_i \lambda_i=1$. In other words, these are convex combinations of vertices $\vec{p}_i$. Each vertex $\vec{p}_i$ consists of all the probabilities of detection events at single, pair, and triple sites such as $P_{v}(A)$, $P_{v}(A,B)$, and $P_{v}(A,B,C)$~\cite{pitowsky2001}, where we omit the setting indices. The dimension of $\mathcal{B}$ is $d=26$ in our scenario of three qubits with two settings per site. The component probabilities of vertices are given by deterministic models, where the joint  probabilities are factorized into the single-site, i.e. $P_v(A, B, C)=P_v(A)P_v(B)P_v(C), P_v(A, B)=P_v(A)P_v(B)$, and the others similarly. Here, the deterministic probabilities $P_v(X)$  are either 0 or 1 for $X=A,B,C$.  Then, the vertices $\vec{p}_i$ have entries 0 and 1 (see Appendix.~\ref{polch}). In other words, the set of deterministic configurations are the vertices of Bell polytope~\cite{Froissart1981} and their number is $2^6$. 

Every facet inequality of the Bell polytope $\mathcal{B}$ is given in a form of
\begin{eqnarray}
\label{facet}
\vec{\mathcal{C}}\cdot\vec{P}\geq \mathcal{C}_0,
\end{eqnarray}
where the lower bound $\mathcal{C}_0$ is a real number and vector $\vec{\mathcal{C}} \in \mathcal{B}$ is the normal vector to the facet hyperplane.
The equality holds for the facet. The facet is identified by $d$ independent vertices $\vec{p}_i$ such that $\vec{\mathcal{C}}\cdot \vec{p}_i = \mathcal{C}_0$, if $\mathcal{C}_0 \ne 0$. If $\mathcal{C}_0=0$, on the other hand, the necessary number of linearly independent vertices drops down to $d-1$, as the null vertex trivially satisfies the equality with all components being zero. We test the linear independence of the vertices by the matrix rank, once the matrix is composed of row vectors with the vertices~\cite{pal2009}.

The tripartite inequality~\eqref{CHM} can be cast in the form of Eq.~\eqref{facet} with the lower bound $\mathcal{C}_0=0$, as shown in Appendix.~\ref{polch}. The equality to the lower bound is satisfied by $2^5$ vertices. Among them we find $d-1$ ($= 25$) linearly independent vertices (see Table~\ref{tabxx} in Appendix.~\ref{polch}). Reminded that the dimension of Bell polytope $\mathcal{B}$ is $d=26$, these imply that {\em geometric tripartite extension of CH inequality~\eqref{CHM} is a facet inequality of the Bell polytope so that it is tight.} With the same function as in the inequality~\eqref{CHM}, we obtain another inequality which is upper bounded by 1, i.e. $\vec{\mathcal{C}}\cdot\vec{P} \le 1$. Among $2^5$ vertices which satisfy the equality to the upper bound, $d$ ($=26$) vertices are found to be linearly independent, implying that this is also a facet inequality of the Bell polytope (see Table~\ref{tabxxx} in Appendix.~\ref{polch}). We get the two tight tripartite inequalities with the given function. This fact is similar to the case of CH inequality, which defines two facets with a single function~\cite{Brunner2014}. On the other hand, the $3$-site CH inequality~\eqref{CHMM} defines a single facet.

\section{Quantum Violation}
\label{section3}
Quantum mechanics  violates the geometric tripartite extension of CH inequality~$\eqref{CHM}$ and $3$-site CH inequalities  $\eqref{CHMM}$ due to the combined effects entanglement of states and measurement incompatibility of observables. Here we illustrate the quantum violations of both the inequalities for GHZ-type~$\eqref{state1}$ and W-type entangled states~$\eqref{state2}.$ Quantum violations are characterized by robustness of violations  and a critical  detection efficiency, below which no quantum violation is obtained.

A three qubit mixed state can be written as
\begin{eqnarray}
\label{state}
\hat{\rho}_v= v|\psi\rangle\langle \psi|+(1-v)\frac{\openone^{\otimes 3}}{8},&
\end{eqnarray}
where $|\psi\rangle$ is a pure three-qubit state and  $v$ is a parameter, called a visibility, with $0 \le v \le 1.$ We consider two classes of three-qubit pure states, e.g,  GHZ state~\cite{GHZ89},
\begin{eqnarray}
\label{state1}
|\text{GHZ}(\alpha)\rangle=\cos\alpha |000\rangle+\sin\alpha|111\rangle,
\end{eqnarray} 
and W state,
\begin{eqnarray}
\label{state2}
&|W(\theta, \phi)\rangle=\sin\theta \cos\phi|001\rangle+ \sin\theta \sin\phi|010\rangle&\nonumber\\& + \cos\theta|100\rangle.&
\end{eqnarray} 
$|\text{GHZ}(\alpha)\rangle$ and $|W(\theta, \phi)\rangle$ are maximally entangled for $\alpha=\frac{\pi}{4},$ and $\theta=\text{arccos}(\frac{1}{\sqrt{3}}),\phi=\frac{\pi}{4},$ respectively.

Three qubits are measured by spatially separated observers. Each of them locally measures with an analyzer, which is oriented alternatively in two different directions. The  measurements are represented by  projectors,
\begin{eqnarray}
\label{obs}
\hat{\Pi}_i^X = \frac{1}{2}(\openone + \vec{b}_i^X \cdot\vec{\sigma}),
\end{eqnarray}
where $\vec{\sigma}=(\hat{\sigma}_x,\hat{\sigma}_y,\hat{\sigma}_z)$ is the vector of Pauli operators $\hat{\sigma}_i$, and Bloch unit vector $\vec{b}_i^X$ stands for $i$-th setting of the analyzer at site $X$. 

When considering Bell inequality in terms of symmetric differences, as in Eq. \eqref{tttt12}, it is convenient to define their positive operators, each by
\begin{eqnarray}
\label{obscp}
\hat{\Pi}^{X \oplus Y} \equiv \hat{\Pi}^X \otimes \openone^Y + \openone^X\otimes\hat{\Pi}^Y - 2 \hat{\Pi}^X \otimes \hat{\Pi}^Y,
\end{eqnarray}
where $\openone^X$ is an identity operator at site $X$, $\hat{\Pi}^X$ a projector as in Eq.~\eqref{obs}, and `$\otimes$' is the tensor product. The symmetric differences $X \oplus Y$ in quantum theory are nominal and their expansions are inapplicable to the set operations. Eq.~\eqref{obscp} can be extended to more sites recursively, for instance, $Y \rightarrow Y \oplus Z$, where $\openone^{Y\oplus Z}$ is set to $\openone^Y \otimes \openone^Z$. Then, those for the pair- and triple-site joint probabilities are given as
\begin{eqnarray}
\label{obscp2}
\hat{\Pi}^{X \oplus Y} &=& {1 \over 2} \left( \openone^{X \oplus Y} - \vec{b}^X\cdot \vec{\sigma} \otimes \vec{b}^Y\cdot \vec{\sigma}\right) \\
\hat{\Pi}^{A \oplus B \oplus C} &=& {1 \over 2} \left( \openone^{A \oplus B \oplus C} + \bigotimes\limits_{X=A,B,C} \vec{b}^X\cdot \vec{\sigma} \right),
\end{eqnarray}
where $\vec{b}^X$ are Bloch vectors of the settings at site $X=A,B,C$. Here we assume the perfect detection.

The quantum joint probability of triple sites, that all the three qubits are  detected for given settings, is given by
\begin{eqnarray}
\label{pro}
P_0(A_i, B_j, C_k) = \text{Tr}\left(\hat{\rho}_v \, \hat{\Pi}^A_i\otimes\hat{\Pi}^B_j\otimes\hat{\Pi}^C_k\right),
\end{eqnarray}
where $\hat{\rho}_v$ is a quantum state, as in Eq.~\eqref{state}. 
Single- and pair-site probabilities are obtained similarly. Here, the detectors together with the analyzers are assumed to work perfectly with unit efficiency $\eta=1$. In realistic situations, however, detectors can be imperfect with less efficiencies than the perfect detectors. Then the quantum probabilities are functions of efficiencies,
\begin{eqnarray}
\label{pro1}
P(X_i) &=& \eta P_0(X_i),  \nonumber\\
P(X_i, Y_j) &=& \eta^2 P_0(X_i, Y_j), \nonumber \\
P(A_i, B_j, C_k) &=& \eta^3 P_0(A_i, B_j, C_k),
\end{eqnarray}  
where $X,Y=A,B,C$ with $Y \ne X$, and $P_0(X_i),$ $P_0(X_i, Y_j),$ $P_0(A_i, B_j, C_k)$ are  quantum probabilities with perfect detections, as in Eq.~\eqref{pro}. Here we assume that the detectors have the same efficiency $\eta$ with $0\le \eta \le 1$. Also, projectors in Eqs.~\eqref{obscp} and \eqref{obscp2} are replaced by $\eta \hat{\Pi}^X$.

We check the violations of the inequality~$\eqref{CHM}$  by replacing Kolmogorov probabilities with the quantum ones, as in Eqs.~\eqref{pro1}. We characterize  quantum violations in terms of experimental imperfections. In particular, two types of imperfections are accounted in measurements and states, i.e., the loss of particles in terms of detection efficiency $\eta$, as in Eq.~\eqref{pro1}, and the depolarization by environment in terms of $1-v,$ as in Eq.~\eqref{state}. Quantum violations can be achieved only if $v > v_{\text{crit}}.$ We find the critical visibility $v_{\text{crit}}$ for a given pure state $|\psi\rangle$ and detection efficiency $\eta=1,$ as the local measurement settings are optimized. For, a given pure state, we define the robustness of violation against the white noise as 
\begin{eqnarray}
\label{eq:vd}
n \equiv 1-v_\text{crit},
\end{eqnarray}
It is also considered  as a violation degree (or the maximal fraction of white noise for which the violations  are not found). For the given pure state $|\psi\rangle,$ we decrease the detection efficiency down to $\eta \le \eta_\text{crit},$ such that no quantum violation can be observed even with the maximal visibility $v=1.$ We also find the minimum of the critical detection efficiencies $\eta_{\text{crit}}^{\text{min}}$  over all possible  pure states. $\eta_{\text{crit}}^{\text{min}}$ is obtained by looking for the lowest eigen value of Bell operator associated to geometric tripartite extension of CH inequality~$\eqref{CHM}.$ Similar kind of analysis was done for CH inequality in Ref.~\cite{eberhard1993}.

\subsection{Critical visibilities and detection efficiencies}
\label{section4}

The Bell operator $\hat{B}$ corresponding to the  inequality~$\eqref{CHM}$ is written as
\begin{eqnarray}
\label{CHM1}
\hat{B} &=& \hat{L}+ \hat{Q}_{AB}+\hat{Q}_{BC}+ \hat{Q}_{CA} + 2\hat{T},
\end{eqnarray}
where $\hat{L}$ is the sum of local (single-site) projectors, $\hat{Q}_{XY}$ is a combination  of pair-site projectors for subsystems $X$ and $Y$, and $\hat{T}$ is a combination of triple-site projectors. Their explicit forms, including the detection efficiency $\eta$, are given by
\begin{eqnarray}
\hat{L} &=& \eta \sum_{X=A,B,C} \hat{\Pi}^X_2 \nonumber \\
\hat{Q}_{XY} &=&\eta^2 \big( \hat{\Pi}^X_1\otimes\hat{\Pi}^Y_1-\hat{\Pi}^X_1\otimes\hat{\Pi}^Y_2-\hat{\Pi}^X_2\otimes\hat{\Pi}^Y_1 \nonumber \\
&& - \hat{\Pi}^X_2 \otimes\hat{\Pi}^Y_2 \big)\nonumber \\
\hat{T} &=& \eta^3 \big(\hat{\Pi}^A_1\otimes\hat{\Pi}^B_2\otimes\hat{\Pi}^C_2+\hat{\Pi}^A_2\otimes\hat{\Pi}^B_1\otimes\hat{\Pi}^C_2 \nonumber \\
&& +\hat{\Pi}^A_2\otimes\hat{\Pi}^B_2\otimes\hat{\Pi}^C_1  - \hat{\Pi}^A_1\otimes\hat{\Pi}^B_1\otimes\hat{\Pi}^C_1 \big).
\end{eqnarray}

The quantum violation of the inequality~$\eqref{CHM}$ is witnessed when $\langle \hat{B} \rangle \equiv \text{Tr}(\hat{\rho}\hat{B}) < 0$. Let $|\psi\rangle$ be a given pure state and $\hat{\rho}_v$ be the mixed state, as in Eq.~\eqref{state}. The critical visibility $v_{\text{crit}}$ is defined by $v$, such that
\begin{eqnarray}
\label{eq:ccv}
\langle \hat{B} \rangle_v \equiv \text{Tr}(\hat{\rho}_{v} \hat{B})=0,
\end{eqnarray}
where $\langle \hat{B} \rangle_v$ is minimized over local measurements settings. Here, $\eta$ is a given detection efficiency.

\subsubsection{Critical visibility for GHZ states} 

\begin{figure}
 \includegraphics[width=0.45\textwidth]{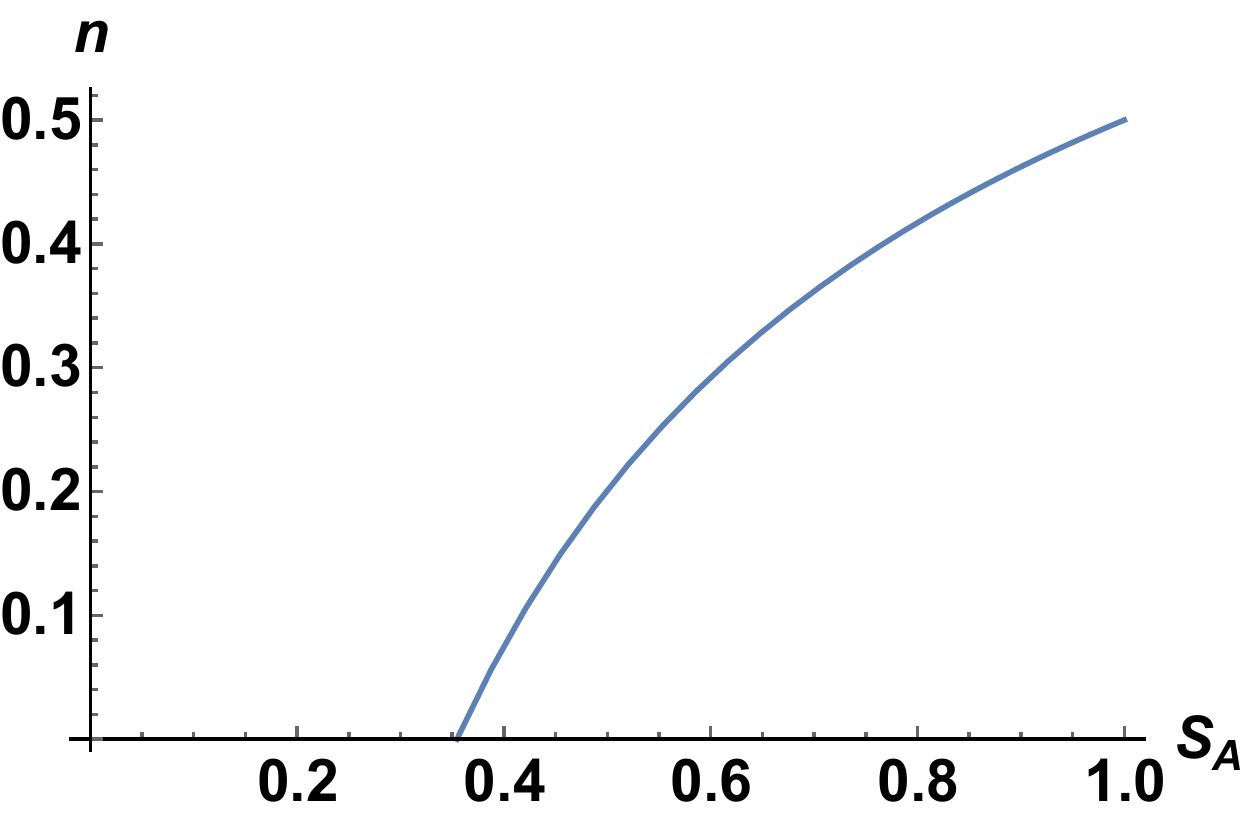}
\caption{Violation robustness $n$ against the white noise, as a function of entanglement $S_A$ for GHZ states with the perfect detectors of $\eta=1$. The robustness $n$ increases as increasing the entanglement $S_A$, whereas no violation is revealed for  partially entangled GHZ states with $\sin2\alpha \le 1/2$. For the maximally entangled GHZ state, i.e., $|\text{GHZ}(\pi/4)\rangle,$ the violation robustness $n= 1/2$.}
\label{fig01}
\end{figure}

At first, we analyze the violations of  the  inequality~$\eqref{CHM}$ by  GHZ states \eqref{state1}. Assuming detectors with perfect efficiency $\eta=1,$ we find the critical visibility $v_\text{crit}$ from Eq.~\eqref{eq:ccv}. Here, $\langle \hat{B} \rangle_v$ is found to be minimized, if the local settings $\vec{b}$ in Eq.~\eqref{pro} are chosen along the directions in the $x$-$y$ plane,
\begin{eqnarray}
\label{obs1}
\vec{b}^X_i=(\cos\theta^X_i, \sin\theta^X_i,0),
\end{eqnarray}
where $\theta^X_i$ are the angles. 
For a GHZ state~\eqref{state1} the critical visibility reads
\begin{eqnarray}
\label{sol1}
v_{\text{crit}}= \min_{\{\theta^X_i \}} {1 \over 1-2\langle \psi | \hat{B} | \psi \rangle} = \frac{2}{M_\text{GHZ}},
\end{eqnarray}
where $M_\text{GHZ} = \max_{\{\theta^X_i\}} \left( \Theta_{111} - \Theta_{122} - \Theta_{212} - \Theta_{221} \right)$, and $\Theta_{ijk}=\langle \text{GHZ} |\,  \vec{b}^A_i \cdot \vec{\sigma} \otimes \vec{b}^B_j \cdot \vec{\sigma} \otimes \vec{b}^C_k \cdot \vec{\sigma}  \, | \text{GHZ} \rangle.$ We find $M_\text{GHZ} = 4 \sin 2\alpha,$ where we  have used $\cos\theta_{122}+\cos\theta_{212}+\cos\theta_{221}-\cos\theta_{111} \le 4$ with $\theta_{ijk}=\theta^A_i + \theta^B_j + \theta^C_k.$ Then, the critical visibility $v_{\text{crit}}$ is given as
\begin{eqnarray}
\label{sol2}
v_{\text{crit}}=\frac{1}{2\sin 2\alpha}.
\end{eqnarray}
From this, we obtain the violation robustness $n$ in Eq.~\eqref{eq:vd},
\begin{eqnarray}
\label{sol2vd}
n=1 - \frac{1}{2\sin 2\alpha}.
\end{eqnarray}
In Fig.~\ref{fig01}, we present the violation robustness $n$ as a function of the entanglement degree $S_A$, where $S_A$ is the von-Neunmann entropy of the reduced density operator $\hat{\rho}_A$ for Alice's qubit, i.e. 
$$
S_A=\sin^2\alpha \, \ln \left( \cos^2\alpha/\sin^2\alpha \right)-\ln \left( \cos^2\alpha \right).
$$ 
This clearly shows that the robustness $n$ increases as increasing the entanglement $S_A,$ whereas no violation is presented for some partially entangled GHZ states with $\sin2\alpha \le 1/2$ (this coincides with the result with Mermin inequality~\cite{scarani2001}). For the maximally entangled GHZ state with $\alpha=\pi/4,$ the violation robustness $n$ is maximal, e.g., $n= 1/2.$ As a reference, for a maximally entangled Einstein-Podolski-Rosen (EPR) state, e.g. $\left( |00\rangle + |11\rangle \right)/\sqrt{2}$, and CH inequality~\cite{Wiesniak2007}, $n = 1- 1/\sqrt{2} \approx 0.2929.$ 

\subsubsection{Critical visibility for W states}
Assuming perfect detection efficiency $\eta=1,$  we check the critical visibility associated to W state \eqref{state2} and the geometric tripartite extension of CH inequality~$\eqref{CHM},$ as done for the GHZ state. For W states, the local settings $\vec{b},$ in Eq.~\eqref{pro}, lie in the $x$-$z$ plane,
\begin{eqnarray}
\label{obs2}
\vec{b}^X_i=(\cos\phi^X_i, 0, \sin\phi^X_i),
\end{eqnarray}
where $\phi^X_i$ are angles. Similarly to Eqs.~\eqref{sol1} and \eqref{sol2}, the critical visibility
\begin{eqnarray}
\label{min1}
v_\text{crit}= \min_{\{\phi^X_i\}} {1 \over 1- 2\langle W | \hat{B} | W \rangle} = {2 \over M_W},
\end{eqnarray}
where $M_W = \max_{\{\phi^X_i\}} \left( \Phi_{111} - \Phi_{122} - \Phi_{212} - \Phi_{221} \right)$ and $\Phi_{ijk}=\langle W |\,  \vec{b}^A_i \cdot \vec{\sigma} \otimes \vec{b}^B_j \cdot \vec{\sigma} \otimes \vec{b}^C_k \cdot \vec{\sigma}  \, | W \rangle$.
For the maximally entangled W state $|W(\text{arccos}(\frac{1}{\sqrt{3}}),\frac{\pi}{4})\rangle $ in Eq.~\eqref{state2}, $M_W \approx 3.0459$~\cite{cabello2002b} so that 
$$
v_{\text{crit}}\approx 0.6566 \quad \text{and} \quad n \approx 0.3434.
$$
The maximum of robustness for the maximally entangled W state is less than $(n=0.5)$ of the maximally entangled GHZ state  but larger than $(n=1-1/\sqrt{2})$  of the maximally entangled EPR state. For other W states, we perform numerical analyses.

\subsubsection{Critical efficiency of detection}
We find the critical efficiency of detection, $\eta_\text{crit}$, above which a pure state $|\psi\rangle$ violates the geometric tripartite extension of CH inequality $\eqref{CHM}.$ For the purpose we employ Bell operator in Eq.~\eqref{CHM1} and investigate its lowest eigenvalue. The critical detection efficiencies for GHZ and W states are to be found by restricting to the subspaces of GHZ and W. Their minima are compared to the critical detection efficiency $\eta_\text{crit}^\text{min}$ to be found on the whole Hilbert space. Here, we shall obtain critical detection efficiencies by assuming particular sets of local settings and numerically confirm them without the assumption.

We consider Bell operator $\hat{B}$ in Eq.~\eqref{CHM1}, assuming that the three observers choose the same alternating local measurements with Bloch vectors $\vec{b}^X_i = \vec{b}_i$ for $X=A,B,C$. Furthermore $\vec{b}_i = (\cos \phi_i, 0, \sin \phi_i)$ lies in the $x$-$z$ plane, where $\phi_1 = \pi/2$ and $\phi_2 = \pi/2 - \phi.$ One might release this constraint for more general settings and obtain the same result. Letting $\hat{B} = \hat{B}(\phi),$ the eigenvalues $\hat{B}(\phi)$ are determined by the roots of the characteristic polynomial $\chi(\lambda, \phi) = \det(\lambda \openone - \hat{B}(\phi)).$ Let the roots $\lambda_r$ be functions of $\phi,$ $\lambda_r=\lambda_r(\phi).$ If $\phi=0$, all the settings are equal to each other and they are compatible, revealing no violation. However, for small $\phi>0$, one finds the lowest eigenvalue of $\hat{B}$, that is negative. In the case we apply a perturbation theory by expanding $\chi(\lambda, \phi)$ and its roots $\lambda_r(\phi)$ in powers of $\phi$ upto $4$th orders. The approximate equation for the roots is given as
\begin{eqnarray}
\label{eq:evalbo}
\chi\left[\lambda_r(\phi),\phi\right] \approx \sum_{n=0}^4 q_n \left(\{ \lambda_{r,m} \} \right) \phi^n = 0,
\end{eqnarray}
where the coefficients $\lambda_{r,n}$ come from the expansion of $\lambda_r(\phi) \approx \sum_{n=0}^4 \lambda_{r,n} \phi^n$. By solving the equations $q_n(\{\lambda_{r,m}\}) = 0$ for all $n$, we obtain $\lambda_{r,n}$ and thus the approximate roots $\lambda_r(\phi)$.

We first find the approximate roots on the subspace, spanned by $\{|001\rangle, |010\rangle, |100\rangle, |111\rangle \}$, which includes W states and EPR states such as $|1\rangle \otimes \left( c_{00} |00\rangle + c_{11}|11\rangle \right).$ The subspace can be called as W-EPR. Note that our tripartite inequality is reduced to a bipartite CH inequality, and with the latter the critical detection efficiency $\eta_\text{crit} \rightarrow 2/3$ in the limit of product states, $c_{11} \rightarrow 1.$ On the W-EPR subspace, the lowest approximate root (i.e., eigenvalue) is given by
\begin{eqnarray}
\label{eq:evalboas}
\lambda_r(\phi) \approx {3\eta \left(2-3 \eta \right) \over 32 \left(1-\eta \right)} \phi^4 < 0, \quad \text{if $\eta > 2/3$}.
\end{eqnarray}
Thus, on the W-EPR subspace, the critical detection efficiency
$$
\eta_\text{crit}={ 2 \over 3 }.
$$ 
 The eigenvector with the lowest eigenvalue is of the form
\begin{eqnarray*}
\label{eigen}
|\psi\rangle=\frac{1}{\sqrt{1+r^2}}(r_1|001\rangle+r_2|010\rangle +r_3|100\rangle +|111\rangle),
\end{eqnarray*}
where $r=\sqrt{r_1^2+r_2^2+r_3^2}.$ The critical detection efciency $\eta_\text{crit}=2/3$ remains unchanged for states with any $r_{1,2,3}$ being non-zero. In other words, $\eta_\text{crit}=2/3$ for W states as well as for EPR states. We apply the similar method to the whole Hilbert space and find the same value of critical detection efficiency, $\eta_\text{crit}^\text{min}=2/3.$ This is equal to that with a bipartite CH-inequality \cite{eberhard1993}.

Some GHZ states do not violate the geometric tripartite extension of CH inequality inequality not apply the previous method. Instead we directly diagonalize Bell operator $\hat{B}$ on the subspace of GHZ states, spanned by $\{|000\rangle, |111\rangle\}.$ Here we assume the settings $\vec{b}_i = (\cos \theta_i, \sin \theta_i, 0)$, where $\theta_1=\theta$ and $\theta_2=\pi/2.$ The settings were chosen in Eq.~\eqref{obs1} to find the critical visibility for the GHZ states. One might release the constraint for more general settings and obtain the same result. The lowest eigenvalue of Bell operator $\hat{B}$ is given by
\begin{eqnarray}
\label{eq:evghz}
\lambda_\text{min} = {3 \over 2} \left(1 - \eta \right) + {1 \over 2} \eta^2 \left( 1 -  \sqrt{1 + 3 \cos^2 \theta } \right).
\end{eqnarray}
It follows from Eq.~\eqref{eq:evghz} that on the GHZ subspace, the critical detection efficiency 
$$
\eta_\text{crit}={\sqrt{21}-3 \over 2},
$$
for which $\theta=0.$ Using Garg-Mermin approach \cite{garg1987}  for Mermin inequality and GHZ state, similar bound for $\eta_\text{crit}$ is obtained in Ref.~\cite{hoban2012}. The respective eigenvector is a maximally entangled GHZ state,
$
|\text{GHZ}(\frac{\pi}{4})\rangle = \left( |000\rangle + |111\rangle \right)/\sqrt{2}.
$ 
The minimum critical detection efficiency for the GHZ states is $0.7913,$ which is larger than $\eta_\text{crit}^\text{min}=\frac{2}{3},$ for W-EPR states.

\subsection{Numerical calculations}

\begin{figure}[tbh]
\includegraphics[scale=0.5]{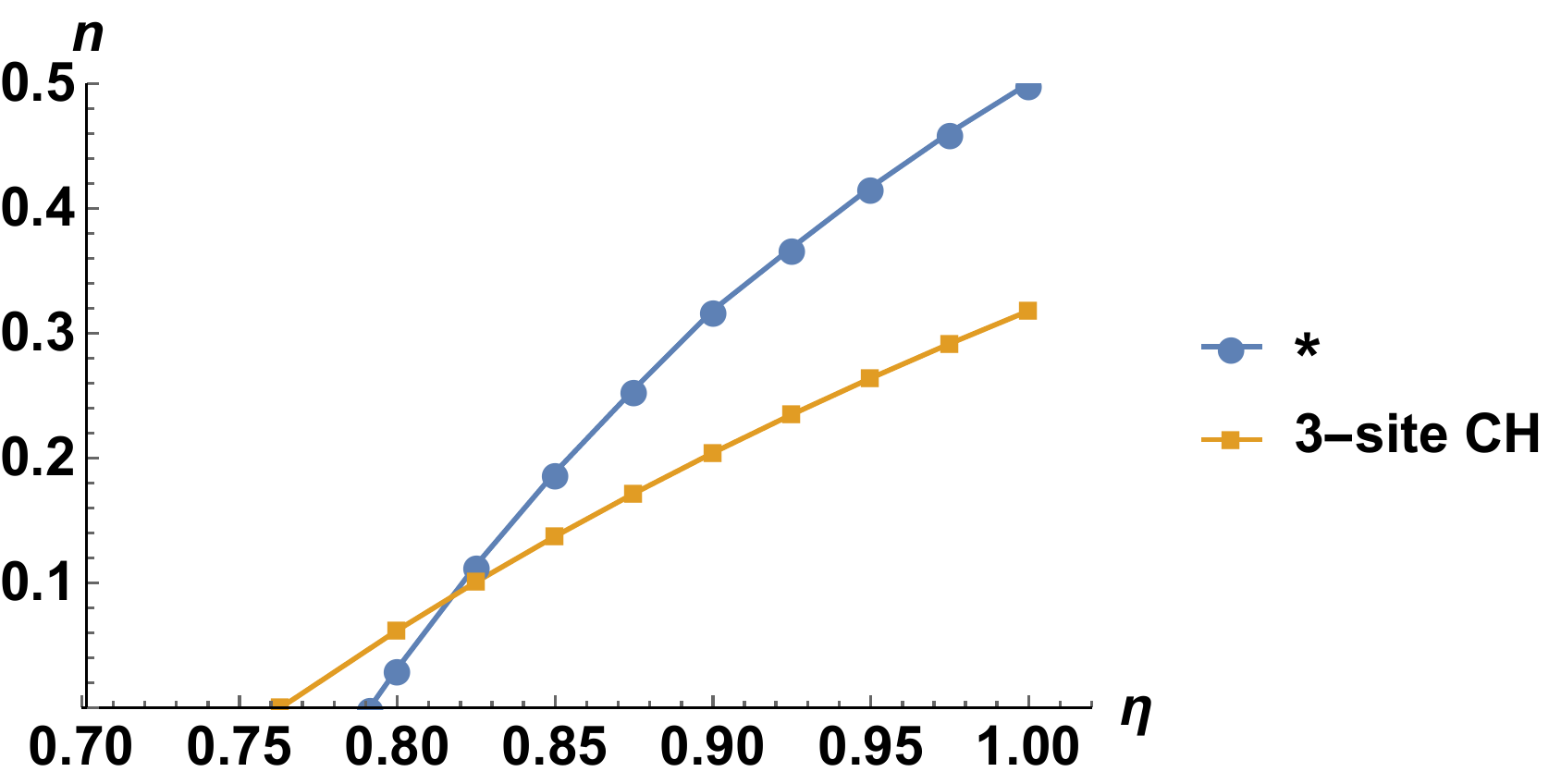}
\caption{Violation robustness $n$ as a function of detection efficiency $\eta$ with the maximally entangled GHZ state $|\text{GHZ}(\frac{\pi}{4})\rangle$ for the geometric tripartite extension of CH inequality \eqref{CHM}* and 3-site CH inequality \eqref{CHMM}, respectively. The inequality~\eqref{CHM} provides larger robustness to white noise than 3-site CH inequality, even though there is a crossover at low $\eta.$ To the contrary, the critical detection efficiency, the minimum $\eta$ to obtain the violation is larger in case of the inequality~\eqref{CHM}.}
\label{fig1}
\end{figure}

%
%
%
%
%

\begin{figure}[tbh]

\includegraphics[scale=0.5]{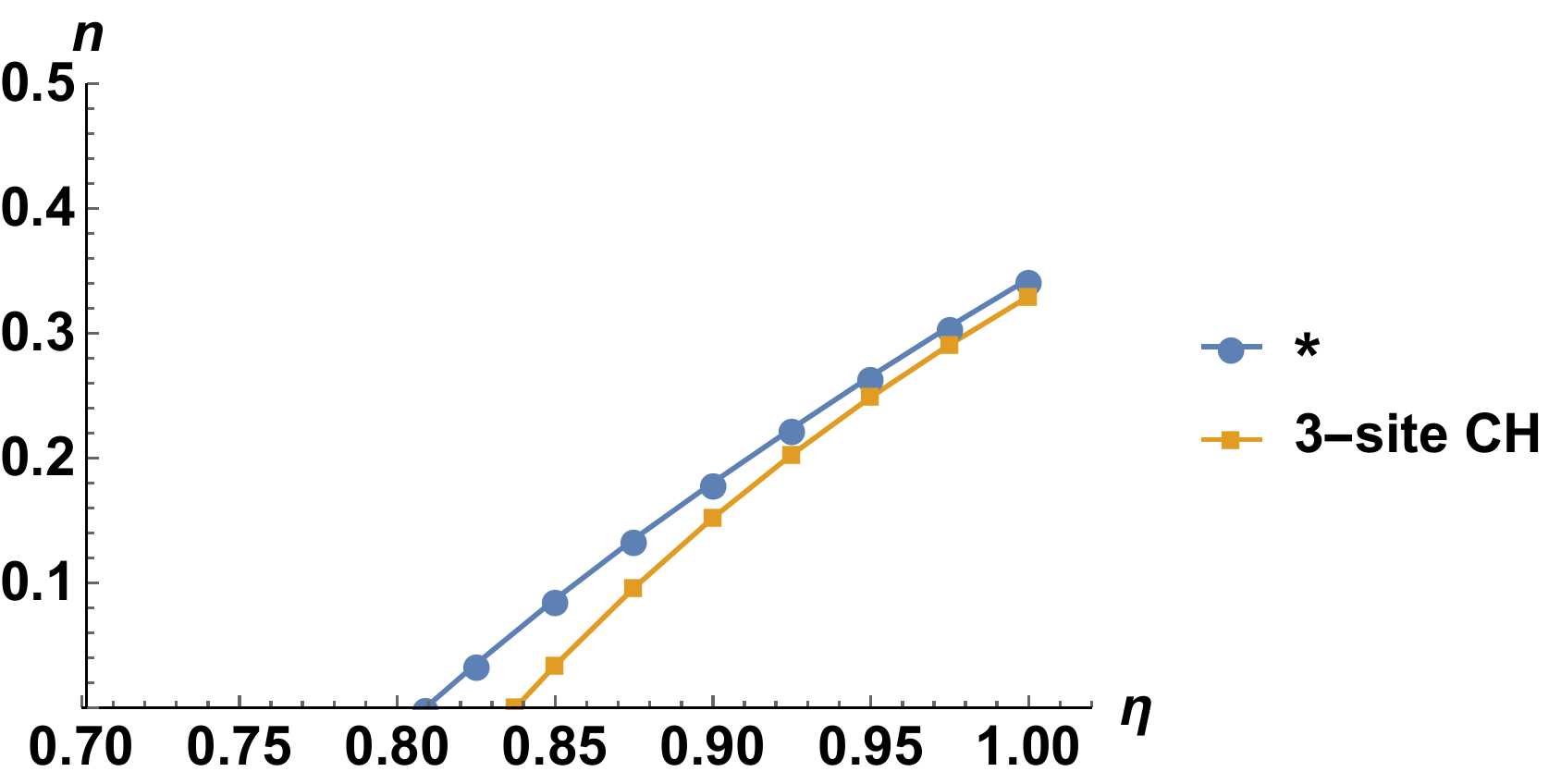}
\caption{With the maximally entangled W state $|\text{W}(\text{arccos}(\frac{1}{\sqrt{3}}),\frac{\pi}{4})\rangle,$ violation robustness $n$ as a function of detection efficiency $\eta$ for the geometric tripartite extension of CH inequality \eqref{CHM}* and 3-site CH inequality \eqref{CHMM}, respectively. The violation robustness is larger in the inequality \eqref{CHM} than 3-site CH inequality over all $\eta.$ The critical detection efficiency is smaller in the geometric tripartite extension of CH inequality, contrary to the GHZ state in Fig.~\ref{fig1}.}
\label{fig3}
\end{figure}

We present our numerical results for the geometric tripartite extension of CH inequality~\eqref{CHM} and 3-site CH inequality~\eqref{CHMM}. In Fig.~\ref{fig1}, we plot the robustness of violation $n$ as a function of detection efficiency $\eta$ for the maximally entangled GHZ state  with the geometric tripartite extension of CH inequality and 3-site CH inequalities. The violation robustness $n$ is a monotonically increasing function of detection efficiency $\eta$ for the maximally entangled GHZ state, while it vanishes if $\eta \le \eta_\text{crit}$. It is shown that the inequality \eqref{CHM} provides larger robustness of violation than 3-site CH inequality, even though there is a crossover at low $\eta.$ To the contrary, the critical detection efficiency, the minimum $\eta$ to see the violation is larger in the geometric tripartite extension of CH inequality inequality. Fig.~\ref{fig3} analyses robustness of violations for the maximally entangled W state. From Fig.~\ref{fig3} we see that for the maximally entangled W state, the violation robustness is larger in the inequality \eqref{CHM} than 3-site CH inequality over all values of $\eta$ between $0$ and $1.$ The critical detection efficiency is smaller in the geometric tripartite extension of CH inequality, contrary to the GHZ state in Fig.~\ref{fig1}. 

Comparing the two states with the geometric tripartite extension of CH inequality in Figs.~\ref{fig1} and \ref{fig3}, it is remarkable that the maximally entangled GHZ state is more robust against white noise than the maximally entangled W state over all values of detection efficiency $\eta$ between $0$ and $1.$ In particular, the critical detection efficiency $\eta_\text{crit} = (\sqrt{21}-3)/2 \approx 0.7913$ for the GHZ state, slightly smaller than $\eta_\text{crit}\approx 0.8090$ for the W state. With the 3-site CH inequality, to the contrary, the maximally entangled GHZ state is less robust than that of W state.

\begin{table}[tbh]
\caption{Violation robustness $n$, critical visibilities $v_\text{crit}$, and critical detection efficiencies $\eta_\text{crit}$ for the geometric tripartite extension of CH inequality \eqref{CHM}* and 3-site CH inequality \eqref{CHMM} with GHZ states \eqref{state1}.}
\centering
\begin{ruledtabular}
\begin{tabular}{l  l  c   c   c   c}
		&				& \multicolumn{4}{c}{$\alpha$} \\
Type		&  				& {$2\pi/18$} 	& {$3\pi/18$} 	& {$4\pi/18$} 	& {$\pi/4$} \\
  \hline
* 	& $n$		       	& 0.2221 		& 0.4227 		& 0.4923 		& 0.5000 \\
		& $v_\text{crit}$       	& 0.7779 		& 0.5773 		& 0.5077 		& 0.5000 \\
		& $\eta_\text{crit}$  	& 0.9195 		& 0.8312 		& 0.7954 		& 0.7913 \\
  \hline
3-site CH 	& $n$		       	& 0.1781 		& 0.2666 		& 0.3121 		& 0.3178 \\
		& $v_\text{crit}$       	& 0.8219 		& 0.7334 		& 0.6879 		& 0.6822 \\
		& $\eta_\text{crit}$  	& 0.7490 		& 0.7588 		& 0.7626 		& 0.7631 \\
\end{tabular}
\end{ruledtabular}
\label{tab1}
\end{table}

\begin{table}[tbh]
\caption{Violation robustness $n$, critical visibilities $v_\text{crit}$, and critical detection efficiencies $\eta_\text{crit}$ for the geometric tripartite extension of CH inequality \eqref{CHM}* and 3-site CH inequality \eqref{CHMM} with W states~\eqref{state2}}
\centering
\begin{ruledtabular}
\begin{tabular}{l  l  c   c   c   c}
		&				& \multicolumn{4}{c}{($\theta,\phi$)} \\
Type		&  				& $({\pi \over 6},{\pi \over 4})$ & $({\pi \over 2},{\pi \over 4})$ & $({\pi \over 4},{\pi \over 4})$ & $({\text{arccos}(\frac{1}{\sqrt{3}}}),{\pi \over 4})$\\
\hline
*	& $n$		& 0.2554       	& 0.2929		& 0.3333				& 0.3434	 \\	
		& $v_\text{crit}$     & 0.7446  	& 0.7071		& 0.6667				& 0.6566	 \\
		& $\eta_\text{crit}$  & 0.7636		& 0.8284		& 0.8024			& 0.8090 \\
\hline
3-site CH 	& $n$	& 0.2488	       	& 0.2929		& 0.3187				& 0.3293	 \\	
		& $v_\text{crit}$     & 0.7512  	& 0.7071		& 0.6813				& 0.6707	 \\
		& $\eta_\text{crit}$  & 0.8004		& 0.8724		& 0.8367			& 0.8375	 \\
\end{tabular}
\end{ruledtabular}
\label{tab3}
\end{table}

For a set of  GHZ states~$\eqref{state1},$ and  both types of inequalties the violation robustnesses $n$, critical visibilities $v_\text{crit},$ and critical detection efficiencies $\eta_{\text{crit}}$ are presented in Table~\ref{tab1}. The analyses are done for $\alpha = 2\pi/18,$ $3\pi/18,$ and $4\pi/18,$ for which the entanglement degree $S_A \approx 0.52,$ $0.81$, and $0.98$, respectively. For $\eta=1,$ the robustnesses of violations $n$ increase for the geometric tripartite extension of CH inequality and 3-site CH inequality as we increase the  degree of entanglement $S_A$ (or $\alpha$). The violations of the inequality \eqref{CHM} have the larger robustnesses than the 3-site CH for all values of $\alpha.$ However, critical detection efficiency $\eta_\text{crit}$ decreases in case of  the  ineaquality \eqref{CHM}, whereas it increases in the 3-site CH inequality. The maximum of  critical detection efficiencies of 3-site CH, $\eta_\text{crit}^\text{max}\approx 0.7631,$ is smaller than the minimum of detection efficiencies of the geometric tripartite extension of CH inequality, $\eta_\text{crit}^\text{min} \approx 0.7913.$ The minimum of  critical detection efficiencies $\eta_\text{crit}^\text{min}$ associated to  3-site CH inequality is estimated to be $0.6$ in the GHZ subspace~\cite{pal2015}. 

Table~\ref{tab3} presents the results for a set of W states~\eqref{state2}. For $\eta=1,$ the robustnesses of the violations of the geometric tripartite extension of CH inequality $n$ are larger than those of the 3-site CH inequality. Furthermore, critical detection efficiencies $\eta_\text{crit}$ for the inequality \eqref{CHM} are lower than those   associated to 3-site CH inequality.
For the biseparable EPR state ($\theta=\frac{\pi}{2}$ and $\phi=\frac{\pi}{4},$ or specifically, $|0\rangle \otimes \left( |01\rangle + |10\rangle \right)/\sqrt{2}$) and  the geometric tripartite extension of CH inequality inequality, the critical detection efficiency $\eta_\text{crit}= 2/(1 + \sqrt{2}) \approx 0.8284,$ which coincides with that of the critical detection efficiency associated to CH inequality and the EPR state~\cite{eberhard1993,Clauser74, Wiesniak2007}. The value of $\eta_\text{crit}$ for 3-site CH inequality and the biseparable EPR state is obtained as $0.8724,$ which is greater than the critical detection efficiency obtained for the geometric tripartite extension of CH inequality.


\section{Extension of CH inequality  for more measurement settings than two per site and more subsystems than three}
\label{ext}

First, we present an extension of CH inequality for three subsystems and three measurement  settings per site. Then, following the similar method we obtain a CH inequality for four qubits with three measurement settings per site. 

Let's consider a set of events $A_l, B_m, C_n$ for Alice, Bob, and Charlie, respectively, where $l, m, n=1,2,3.$ For detailed descriptions, the reader is referred to subsection~\ref{chmm}. As discussed in case of three subsystems and two measurement settings per site, we can find similar kind of relations between symmetric differences of three events for the situation at hand. For example, for a pair of statistical separations, the following inequality holds: 
\begin{eqnarray}
\label{three1}
&P\big((A_1\oplus B_2)\oplus C_3\big) + P\big((A_3\oplus B_3)\oplus C_3\big)&\nonumber\\ &\geq P\big((A_1\oplus B_2)\oplus (B_3\oplus A_3) \big)&\nonumber\\&=P(A_1\oplus B_2\oplus B_3\oplus A_3).& 
\end{eqnarray}
Also, due to permutation symmetry, $P(A_1\oplus B_2\oplus B_3\oplus A_3)=P\big((A_1\oplus B_2\oplus B_3)\oplus A_3 \big).$ Using the fact, we can have another triangle inequality 
\begin{eqnarray}
\label{three2}
&P\big((A_1\oplus B_2\oplus B_3)\oplus A_3 \big) + P\big(A_3\oplus(B_1\oplus C_2)\big) &\nonumber\\& \geq P\big((A_1\oplus B_1\oplus B_3)\oplus (B_2\oplus C_2)\big)&\nonumber\\&=P(A_1\oplus B_1\oplus B_3\oplus B_2\oplus C_2).& 
\end{eqnarray}
Following the similar argument, we obtain
\begin{eqnarray}
\label{three3}
&P\big((A_1\oplus B_1\oplus B_3)\oplus (B_2 \oplus C_2) \big) + P\big( A_2 \oplus (B_2\oplus C_2)\big)&\nonumber\\ &\geq P\big((A_1\oplus B_1)\oplus (A_2\oplus B_3)\big)&\nonumber\\&=P(A_1\oplus B_1 \oplus A_2 \oplus B_3),&
\end{eqnarray}
and finally,
\begin{eqnarray}
\label{three4}
&P\big((A_1\oplus B_1)\oplus (A_2\oplus B_3) \big) + P\big((A_2\oplus B_3)\oplus C_1\big) &\nonumber\\
&\geq P\big((A_1\oplus B_1)\oplus C_1\big).&
\end{eqnarray}

With the four inequalities:~$\eqref{three1},$ $\eqref{three2},$ $\eqref{three3},$ and $\eqref{three4}$, a CH inequality for three qubits and three settings per site reads
\begin{eqnarray}
\label{three5}
&P(A_1\oplus B_2\oplus C_3)+ P(A_3\oplus B_3\oplus C_3)&\nonumber\\ &+ P(A_3\oplus B_1\oplus C_2)+P(A_2\oplus B_2\oplus C_2)& \nonumber \\ &+ P(A_2\oplus B_3\oplus C_1)\geq P(A_1\oplus B_1\oplus C_1).&
\end{eqnarray}
It is worth to mention that we can have several invariant forms of the inequality~$\eqref{three5}$ by choosing different sets of symmetric differences associated with statistical separations. 

Let's consider a set of events $A_l, B_m, C_n, D_p $ for Alice, Bob, Charlie and Daniel, respectively, where $l, m, n, p=1,2, 3.$ From earlier discussions we know that the statistical separations obey triangle inequalities. Thus following set of inequalities can be obtained by chaining
\begin{eqnarray}
\label{four1}
&P\big((A_1\oplus (B_1\oplus C_1\oplus D_1)\big)+P\big((A_1\oplus (B_2\oplus C_3\oplus D_3)\big)&\nonumber\\
&\geq P\big((B_1\oplus C_1\oplus D_1)\oplus(B_2\oplus C_3\oplus D_3)\big)&\nonumber\\&=P(B_1\oplus C_1\oplus D_1\oplus B_2\oplus C_3\oplus D_3),&
\end{eqnarray}
using permutation symmetry of symmetric difference, we also have
\begin{eqnarray}
\label{four2}
&P\big(( C_1\oplus D_1\oplus B_2\oplus C_3)\oplus(B_1\oplus D_3)\big)&\nonumber\\
&+P\big((A_3\oplus C_2)\oplus (B_1\oplus D_3)\big)&\nonumber\\
&\geq P\big(( C_1\oplus D_1\oplus B_2\oplus C_3)\oplus (A_3\oplus C_2)\big)&\nonumber\\&=P( C_1\oplus D_1\oplus B_2\oplus C_3\oplus A_3\oplus C_2),&
\end{eqnarray}
next,
\begin{eqnarray}
\label{four3}
 &P\big((A_3\oplus C_1)\oplus ( D_1\oplus B_2\oplus C_2\oplus C_3)\big)&\nonumber\\
 &+P\big((A_3\oplus C_1) \oplus( B_3\oplus D_2)\big)&\nonumber\\
&\geq P\big(( D_1\oplus B_2\oplus C_2)\oplus C_3)\oplus (B_3\oplus D_2)\big)&\nonumber\\&=P(D_1\oplus B_2\oplus C_2\oplus C_3\oplus B_3\oplus D_2),&
\end{eqnarray}
 and finally,
\begin{eqnarray}
\label{four4}
 &P\big((B_3\oplus C_3\oplus D_1)\oplus ( D_2\oplus B_2\oplus C_2)\big)&\nonumber\\& +
P\big((B_3\oplus C_3\oplus D_1) \oplus A_2\big)&\nonumber\\
&\geq P\big(( D_2\oplus B_2\oplus C_2)\oplus A_2)\big).&
\end{eqnarray}
From these four inequalities ~$\eqref{four1},$ $\eqref{four2},$ $\eqref{four3},$ and $\eqref{four4}$ we obtain 
\begin{eqnarray}
\label{four5}
P(A_1\oplus B_1 \oplus C_1 \oplus D_1) + P(A_1\oplus B_2\oplus C_3 \oplus D_3)\nonumber\\
+P(A_3\oplus B_1\oplus C_2 \oplus D_3)+  P(A_3\oplus B_3 \oplus C_1\oplus D_2)\nonumber\\
+P(A_2\oplus B_3 \oplus C_3 \oplus D_1)\geq P( A_2\oplus  D_2\oplus B_2\oplus C_2).
\end{eqnarray}
Quantum mechanics violates the inequalities $\eqref{three5},$ and $\eqref{four5}.$ It is interesting to note that because the basic formalism leading to the  inequalities $\eqref{tttt12},$ $\eqref{three5}$, and $\eqref{four5}$ is same, in principle, we may obtain several CH inequalities for arbitrary number of subsystems with many measurement settings per site. It will be discussed elsewhere.

\section{Remarks}
\label{section5}
We propose a geometric method to construct a set of  CH inequalities in Kolmogorov theory of probability. We show that the  inequalities results from a set of triangle inequalities in terms of the statistical separations, defined by the probability of symmetric differences between local detection events. The method can be generalized to obtain inequalities for more qubits than three  and more measurement settings than two per site. Also, a new set of  inequalities, which lead to Mermin inequalities, are presented (see Appendix.~\ref{CH-Mermin}) upto five subsystems. It is also shown that our  inequality \eqref{CHM} can be reduced to  the CH inequality of two subsystems and the Mermin inequality of 3 qubits under the appropriate conditions.

The inequalities are characterized in terms of  two parameters, e.g., detection efficiency and robustness of quantum violation against white-noise. The later is employed as a degree of violation. For the three qubits, we show that the geometric tripartite extension of CH inequality can be quantum-mechanically violated by the pure entangled states of GHZ and W types, even though no violation is found for some partially entangled GHZ states. In terms of the violation robustness and critical efficiency of detection, we compare our inequality to the 3-site CH inequality \eqref{CHMM}, another generalization of CH inequalities for tripartite systems. Our  inequality \eqref{CHM} shows violation robustness stronger than the 3-site CH inequality by both types of GHZ and W states with the perfect detections. We find that, for the given GHZ states, the 3-site CH inequality has the critical efficiencies lower than the geometric tripartite extension of CH inequality, whereas it is opposite for the given W states. In particular, the minimal critical efficiency of detection over all the states is lower in the 3-site than the the geometric tripartite extension of CH inequality. It is worth to mention that the authors in Ref.~\cite{larsson2001} did not account background noises, even though they conjectured that the inclusion of back ground noise would increase the bounds on the minimum detection efficiency. We find that bi-separable states can violate the inequality \eqref{CHM}, as expected by the fact that two qubit CH inequality can be retrieved from the geometric tripartite extension of CH inequality \eqref{CHM} and violated by two-qubit pure entangled states.

The quantum violations of the geometric tripartite extension of CH inequality \eqref{CHM} are expected to experimentally realize with detectors, which differentiate single- and multi-photon detections with high efficiency and negligible dark count \cite{lita2008}. It is conjectured that the  method to extend CH inequality can be further developed for arbitrary number of outcomes. This will be presented in a forthcoming work.

\section*{ACKNOWLEDGMENTS}
This work is supported by the grant (No. 2014R1A2A1A10050117), funded by National Research Foundation of Korea (NRF) and the Korean government (MSIP). MZ is supported by a DFG-FNP award-grant COPERNICUS.

\appendix
\section{Upper bound of left-hand side of $\eqref{CHM}$}
\label{bound}

Let $[a]$ denote non-negative residue $a$ modulo $2$, 
\begin{eqnarray}
[a] \equiv a \mod 2.
\end{eqnarray}
Two residues $[a]$ and $[b]$ satisfy triangle inequality,
 \begin{eqnarray}
 \label{app0}
 [a+b] \le [a] + [b].
 \end{eqnarray}
Also one may have an identity,
$$[a] = 1-[1- a].$$
Thus, one obtains the relation,
\begin{eqnarray}
\label{app1}
 [a]+[b] \le 2 - [a+ b].
\end{eqnarray}

Using Eqs.~\eqref{KOL1} and \eqref{app1}, we have
\begin{eqnarray}
\label{app2}
P(A_1\oplus B_2\oplus C_2)+ P(A_2\oplus B_1\oplus C_2) \nonumber\\
\leq 2-P(A_1\oplus B_2\oplus A_2\oplus B_1 ).
\end{eqnarray}
Using triangle inequality $\eqref{app0}$, we also obtain
\begin{eqnarray}
\label{app3}
P(A_2\oplus B_2\oplus C_1)\leq P(A_1\oplus B_2\oplus A_2\oplus B_1 )\nonumber\\
+P(A_1\oplus B_1\oplus C_1).
\end{eqnarray}
Summing $\eqref{app2}$ and $\eqref{app3}$ we get
\begin{eqnarray}
\label{app4}
2 \ge&& P(A_1\oplus B_2\oplus C_2)+ P(A_2\oplus B_1\oplus C_2) \nonumber\\
&& +P(A_2\oplus B_2\oplus C_1)-P(A_1\oplus B_1\oplus C_1).
\end{eqnarray}
Combining Eq.~\eqref{app4} and Eq.~\eqref{tttt12}, we obtain
\begin{eqnarray}
\label{app5}
0 \le & P(A_1\oplus B_2\oplus C_2)+ P(A_2\oplus B_1\oplus C_2)  & \nonumber\\
& + P(A_2\oplus B_2\oplus C_1)-P(A_1\oplus B_1\oplus C_1) & \le 2.
\end{eqnarray}
Or equivalently, in the form of Eq.~\eqref{CHM},
\begin{eqnarray}
0 \le L + Q_{AB} + Q_{BC} + Q_{CA} + 2T \le 1.
\end{eqnarray}
Thus we have two bounds, 0 for the lower bound and 1 for the upper bound.
\pagebreak
\begin{widetext}
\section{Polytope for the geometric tripartite extension of CH inequality}
\label{polch}
Vertices $\vec{p}_i$ of the Bell polytope $\mathcal{B}$ are defined,
\begin{eqnarray*}
\vec{p}_i&=&\big( P_v(A_1), P_v(A_2), P_v(B_1), P_v(B_2), P_v(C_1), P_v(C_2), \\
&& P_v(A_1)P_v(B_1), P_v(A_1)P_v(B_2), P_v(A_2)P_v(B_1), P_v(A_2)P_v(B_2),\\
&& P_v(B_1)P_v(C_1), P_v(B_1)P_v(C_2), P_v(B_2)P_v(C_1),  P_v(B_2)P_v(C_2), \\
&& P_v(A_1)P_v(C_1), P_v(A_1)P_v(C_2), P_v(A_2)P_v(C_1),  P_v(A_2)P_v(C_2), \\
&& P_v(A_1)P_v(B_1)P_v(C_1), P_v(A_1)P_v(B_1)P_v(C_2), P_v(A_1)P_v(B_2)P_v(C_1), P_v(A_2)P_v(B_1)P_v(C_1), \nonumber\\
&& P_v(A_1)P_v(B_2)P_v(C_2), P_v(A_2)P_v(B_1)P_v(C_2), P_v(A_2)P_v(B_2)P_v(C_1),P_v(A_2)P_v(B_2)P_v(C_2) \big),
\end{eqnarray*}
where the probabilities $P_v(X)$ are either 0 or 1 for $X=A_i,B_j,C_k$. The vector $\vec{P} = \sum_i \lambda_i \vec{p}_i$ in $\mathcal{B}$ is given in a form of
\begin{eqnarray*}
\vec{P}&=& \big(P(A_1), P(A_2), P(B_1), P(B_2), P(C_1), P(C_2), \\
&& P(A_1, B_1), P(A_1, B_2), P(A_2, B_1), P(A_2, B_2), \\ 
&& P(B_1, C_1),P(B_1, C_2), P(B_2, C_1), P(B_2, C_2), \\
&& P(A_1, C_1), P(A_1, C_2),  P(A_2, C_1), P(A_2, C_2), \\
&& P(A_1, B_1, C_1), P(A_1, B_1, C_2), P(A_1, B_2, C_1), P(A_2, B_1, C_1), \\
&& P(A_1, B_2, C_2), P(A_2, B_1, C_2), P(A_2, B_2, C_1), P(A_2, B_2, C_2)\big).
\end{eqnarray*}
The geometric tripartite extension of CH inequality in Eq.~\eqref{facet} is specified by $\vec{\mathcal{C}}\cdot \vec{P} \ge \mathcal{C}_0$ with $\mathcal{C}_0=0$ and
\begin{eqnarray*}
\vec{\mathcal{C}} = (0, 1, 0, 1, 0, 1, 1, -1, -1, -1, 1, -1, -1, -1, 1, -1, -1, -1,  -2, 0, 0, 0, 2, 2, 2, 0).
\end{eqnarray*}
The inequality \eqref{CHM} is a facet inequality as shown in the main text and the facet is specified by the set of linearly independent vertices. We present them in Table~\ref{tabxx} and also those of another facet inequality with the upper bound 1 in Table~\ref{tabxxx}. 
\end{widetext}

\begin{table}[h]
\caption{Row vectors of vertices in the Bell polytope, which yield the lower bound of the  inequality \eqref{CHM}. These $25$ extreme vertices are linearly independent (excluding the null vector of vertex).}
\centering
\begin{tabular}{c c c c c c c c c c c c c c c c c c c c c c c c c c c c c c c c}
  \hline\hline
  0 & 0 & 0& 0& 1& 0& 0& 0& 0& 0& 0& 0 & 0& 0& 0& 0& 0&0& 0& 0& 0& 0& 0&
   0& 0& 0\\
  0 & 0 & 0& 1& 1& 0& 0& 0& 0& 0& 0& 0 & 1& 0& 0& 0& 0&0& 0& 0& 0& 0& 0&
   0& 0& 0\\
  0 & 0 & 0& 1& 1& 1& 0& 0& 0& 0& 0& 0 & 1& 1& 0& 0& 0&0& 0& 0& 0& 0& 0&
   0& 0& 0\\
  0 & 0 & 1& 0& 0& 0& 0& 0& 0& 0& 0& 0 & 0& 0& 0& 0& 0&0& 0& 0& 0& 0& 0&
   0& 0& 0\\
  0 & 0 & 1& 0& 0& 1& 0& 0& 0& 0& 0& 1 & 0& 0& 0& 0& 0&0& 0& 0& 0& 0& 0&
   0& 0& 0\\
 0 & 0 & 1& 1& 0& 1& 0& 0& 0& 0& 0& 1 & 0& 1& 0& 0& 0&0& 0& 0& 0& 0& 0&
   0& 0& 0\\
  0 & 0 & 1& 1& 1& 1& 0& 0& 0& 0& 1& 1 & 1& 1& 0& 0& 0&0& 0& 0& 0& 0& 0&
   0& 0& 0\\
 0 & 1 & 0& 0& 1& 0& 0& 0& 0& 0& 0& 0 & 0& 0& 0& 0& 1&0& 0& 0& 0& 0& 0&
   0& 0& 0\\
  0 & 1 & 0& 0& 1& 1& 0& 0& 0& 0& 0& 0 & 0& 0& 0& 0& 1& 1& 0& 0& 0& 0& 0&
   0& 0& 0\\
  0 & 1 & 0& 1& 0& 1& 0& 0& 0& 1& 0& 0 & 0& 1& 0& 0& 0&1& 0& 0& 0& 0& 0&
   0& 0& 1\\
 0 & 1 & 0& 1& 1& 1& 0& 0& 0& 1& 0& 0 & 1& 1& 0& 0& 1&1& 0& 0& 0& 0& 0&
   0& 1& 1\\
 0 & 1 & 1& 0& 0& 0& 0& 0& 1& 0& 0& 0 & 0& 0& 0& 0& 0&0& 0& 0& 0& 0& 0&
   0& 0& 0\\
  0 & 1 & 1& 0& 1& 0& 0& 0& 1& 0& 1& 0 & 0& 0& 0& 0& 1&0& 0& 0& 0& 1& 0&
   0& 0& 0\\
  0 & 1 & 1& 1& 0& 0& 0& 0& 1& 1& 0& 0 & 0& 0& 0& 0& 0&0& 0& 0& 0& 0& 0&
   0& 0& 0\\
  0 & 1 & 1& 1& 0& 1& 0& 0& 1& 1& 0& 1 & 0& 1& 0& 0& 0&1& 0& 0& 0& 0& 0&
   1& 0&1\\
  1 & 0 & 0& 0& 0& 0& 0& 0& 0& 0& 0& 0 & 0& 0& 0& 0& 0&0& 0& 0& 0& 0& 0&
   0& 0& 0\\
  1 & 0 & 0& 0& 0& 1& 0& 0& 0& 0& 0& 0 & 0& 0& 0& 1& 0&0& 0& 0& 0& 0& 0&
   0& 0& 0\\
  1 & 0 & 0& 1& 0& 0& 0& 1& 0& 0& 0& 0 & 0& 0& 0& 0& 0&0& 0& 0& 0& 0& 0&
   0& 0& 0\\
  1 & 0 & 0& 1& 1& 0& 0& 1& 0& 0& 0& 0 & 1& 0& 1& 0& 0&0& 0& 0& 1& 0& 0&
   0& 0& 0\\
 1 & 0 & 1& 0& 0& 1& 1& 0& 0& 0& 0& 1 & 0& 0& 0& 1& 0&0& 0& 1& 0& 0& 0&
   0& 0& 0\\
1 & 0 & 1& 0& 1& 1& 1& 0& 0& 0& 1&1& 0& 0& 1& 1& 0&0& 1& 1& 0& 0& 0&
   0& 0& 0\\
  1 & 0 & 1& 1& 1& 0& 1& 1& 0& 0& 1& 0 & 1& 0& 1& 0& 0&0& 1& 0& 1& 0& 0&
   0& 0& 0\\
  1 & 0 & 1& 1& 1& 1& 1& 1& 0& 0& 1& 1 & 1& 1& 1& 1& 0&0& 1& 1& 1& 0& 1&
   0& 0& 0\\
  1 & 1 & 0& 0& 0& 1& 0& 0& 0& 0& 0& 0 & 0& 0& 0& 1& 0&1& 0& 0& 0& 0& 0&
   0& 0& 0\\
  1 & 1 & 0& 0& 1& 1& 0& 0& 0& 0& 0& 0 & 0& 0& 1& 1& 1&1& 0& 0& 0& 0& 0&
   0& 0& 0\\
   \hline
\end{tabular}
\label{tabxx}
\end{table}
\begin{table}[h]
\caption{$26$ linearly independent row vectors of extreme vertices in the Bell polytope, which yield the upper bound}
\centering
\begin{tabular}{c c c c c c c c c c c c c c c c c c c c c c c c c c c c c c c c}
  \hline\hline
0 & 0 & 0& 0& 0& 1& 0& 0& 0& 0& 0& 0 & 0& 0& 0& 0& 0&0& 0& 0& 0& 0& 0&
   0& 0& 0\\
  0 & 0 & 0& 0& 1& 1& 0& 0& 0& 0& 0& 0 & 0& 0& 0& 0& 0&0& 0& 0& 0& 0& 0&
   0& 0& 0\\
  0 & 0 & 0& 1& 0& 0& 0& 0& 0& 0& 0& 0 & 0& 0& 0& 0& 0&0& 0& 0& 0& 0& 0&
   0& 0& 0\\
  0 & 0 & 0& 1& 0& 1& 0& 0& 0& 0& 0& 0 & 0& 1& 0& 0& 0&0& 0& 0& 0& 0& 0&
   0& 0& 0\\
  0 & 0 & 1& 0& 1& 0& 0& 0& 0& 0& 1& 0 & 0& 0& 0& 0& 0&0& 0& 0& 0& 0& 0&
   0& 0& 0\\
  0 & 0 & 1& 0& 1& 1& 0& 0& 0& 0& 1& 1 & 0& 0& 0& 0& 0&0& 0& 0& 0& 0& 0&
   0& 0& 0\\
 0 & 0 & 1& 1& 0& 0& 0& 0& 0& 0& 0& 0 & 0& 0& 0& 0& 0&0& 0& 0& 0& 0& 0&
   0& 0& 0\\
  0 & 0 & 1& 1& 1& 0& 0& 0& 0& 0& 1& 0 & 1& 0& 0& 0& 0&0& 0& 0& 0& 0& 0&
   0& 0& 0\\
 0 & 1 & 0& 0& 0& 0& 0& 0& 0& 0& 0& 0 & 0& 0& 0& 0& 0&0& 0& 0& 0& 0& 0&
   0& 0& 0\\
  0 & 1 & 0& 0& 0& 1& 0& 0& 0& 0& 0& 0 & 0& 0& 0& 0& 0& 1& 0& 0& 0& 0& 0&
   0& 0& 0\\
  0 & 1 & 0& 1& 1& 0& 0& 0& 0& 1& 0& 0 & 1& 0& 0& 0& 1&0& 0& 0& 0& 0& 0&
   0& 1& 0\\
 0 & 1 & 0& 1& 0& 0& 0& 0& 0& 1& 0& 0 & 0& 0& 0& 0& 0&0& 0& 0& 0& 0& 0&
   0& 0& 0\\
 0 & 1 & 1& 0& 0& 1& 0& 0& 1& 0& 0& 1 & 0& 0& 0& 0& 0&1& 0& 0& 0& 0& 0&
   1& 0& 0\\
  0 & 1 & 1& 0& 1& 1& 0& 0& 1& 0& 1& 1 & 0& 0& 0& 0& 1&1& 0& 0& 0& 1& 0&
   1& 0& 0\\
  0 & 1 & 1& 1& 1& 0& 0& 0& 1& 1& 1& 0 & 1& 0& 0& 0& 1&0& 0& 0& 0& 1& 0&
   0& 1& 0\\
  0 & 1 & 1& 1& 1& 1& 0& 0& 1& 1& 1& 1 & 1& 1& 0& 0& 1&1& 0& 0& 0& 1& 0&
   1& 1&1\\
  1 & 0 & 0& 0& 1& 0& 0& 0& 0& 0& 0& 0 & 0& 0& 1& 0& 0&0& 0& 0& 0& 0& 0&
   0& 0& 0\\
  1 & 0 & 0& 0& 1& 1& 0& 0& 0& 0& 0& 0 & 0& 0& 1& 1& 0&0& 0& 0& 0& 0& 0&
   0& 0& 0\\
  1 & 0 & 0& 1& 1& 1& 0& 1& 0& 0& 0& 0 &1& 1& 1& 1& 0&0& 0& 0& 1& 0& 1&
   0& 0& 0\\
  1 & 0 & 0& 1& 0& 1& 0& 1& 0& 0& 0& 0 & 0& 1& 0& 1& 0&0& 0& 0& 0& 0& 1&
   0& 0& 0\\
 1 & 0 & 1& 0& 0& 0& 1& 0& 0& 0& 0& 0 & 0& 0& 0& 0& 0&0& 0& 0& 0& 0& 0&
   0& 0& 0\\
1 & 0 & 1& 0& 1& 0& 1& 0& 0& 0& 1&0& 0& 0& 1& 0& 0&0& 1& 0& 0& 0& 0&
   0& 0& 0\\
  1 & 0 & 1& 1& 0& 0& 1& 1& 0& 0& 0& 0 & 0& 0& 0& 0& 0&0& 0& 0& 0& 0& 0&
   0& 0& 0\\
  1 & 0 & 1& 1& 0& 1& 1& 1& 0& 0& 0& 1 & 0& 1& 0& 1& 0&0& 0& 1& 0& 0& 1&
   0& 0& 0\\
  1 & 1 & 0& 0& 0& 0& 0& 0& 0& 0& 0& 0 & 0& 0& 0& 0& 0&1& 0& 0& 0& 0& 0&
   0& 0& 0\\
  1 & 1 & 0& 0& 1& 0& 0& 0& 0& 0& 0& 0 & 0& 0& 1& 0& 1&0& 0& 0& 0& 0& 0&
   0& 0& 0\\
   \hline
\end{tabular}
\label{tabxxx}
\end{table}


\section{Explicit inclusion of non-detection events}
\subsection{Clauser-Horne and Eberhard inequalities}
\label{sec:eche}

Assume each observer gets 3 outcomes $\{+,-,u\}$, where `$u$' is a undetected outcome. CH inequality is given by
\begin{eqnarray}
\label{eq:chineq1}
P(A_1^+)+P(B_1^+) - P(A^+_1, B^+_1) - P(A^+_1, B^+_2) & \nonumber \\
- P(A^+_2, B^+_1) + P(A^+_2, B^+_2) & \ge 0, \nonumber
\end{eqnarray}
where $P(A^+)$ are probabilities of Alice's detecting the outcome `+' and $P(A^+,B^+)$ are joint probabilities of both detecting outcomes `+.' 
We may expand the single-site probabilities in a form of the pair-site probabilities, 
\begin{eqnarray}
\label{eq:pse1}
P(X^+_1) = \sum_{y=\pm,u} P(X^+_1, Y^y_2).
\end{eqnarray}
CH inequality in Eq.~\eqref{eq:chineq1} then becomes
\begin{eqnarray}
\label{eq:chineq2}
P(A_1^+, B_2^-) + P(A_1^+, B_2^u) & \nonumber \\
+ P(A_2^-, B^+_1) + P(A_2^u, B^+_1) & \nonumber \\
+ P(A^+_2, B^+_2) - P(A^+_1, B^+_1) & \ge 0.
\end{eqnarray}
This version of CH inequality is equivalent to Eberhard inquality~\cite{khrennikov2014a}. 

Consider all the possible data of $N$ events in each pair of local settings, assuming that arbitrarily large $N$ is the same for the 4 pairs of local settings. Then, the CH inequality in terms of probabilities can be changed to the one with the coincident counts,
\begin{eqnarray}
\label{eq:ebineq}
N(A_1^+, B_2^-) + N(A_1^+, B_2^u) & \nonumber \\
+ N(A_2^-, B^+_1) + N(A_2^u, B^+_1) & \nonumber \\
+ N(A^+_2, B^+_2) - N(A^+_1, B^+_1) & \ge 0.
\end{eqnarray}
where $N(A^a_i, B^b_j) = N P(A^a_i, B^b_j)$, the number of coincident counts at both sites with outcome pair ($a,b$) for a given setting pair $(i,j)$. Note that we also account the number of counts at one site and no count at the other site with $N(A^{a \ne u}_i,  B^{b=u}_j)$ or $N(A^{a=u}_i, B^{b \ne u}_j)$. This is an Eberhard inequality.


\subsection{Eberhard inequality for 3 qubits}
\label{sec:eche1}
We derive a 3-qubit Eberhard inequality from our inequality \eqref{CHM}, based on the method in Sec.~\ref{sec:eche}. The inequality in Eq.~\eqref{CHM} is recalled,
\begin{eqnarray*}
 L + Q_{AB} + Q_{BC} + Q_{CA} + 2T \geq 0,
\end{eqnarray*}
where 
\begin{eqnarray*}
L &=& \sum_{X=A,B,C} P(X_2^+) \nonumber \\
Q_{XY} &=& P(X_1^+,Y_1^+) - P(X_1^+, Y_2^+) \\
&& -P(X_2^+,Y_1^+) - P(X_2^+,Y_2^+) \nonumber \\
T&=& P(A_1^+, B_2^+, C_2^+)+ P(A_2^+, B_1^+, C_2^+) \\
&& + P(A_2^+, B_2^+, C_1^+) -P(A_1^+, B_1^+, C_1^+).
\end{eqnarray*}

Similarly to Eq.~\eqref{eq:pse1}, we expand the local- and pair-site probabilities in a form of the triple-site. 
The single-site term $L$ is rewritten by
\begin{eqnarray*}
L = \sum_{X=A,B,C} \sum_{y,z=\pm,u} P\left( X^+_2, Y^y_2, Z^z_1 \right),
\end{eqnarray*}
where $(X,Y,Z)$ is one of cyclic permutations of $(A,B,C)$.
The pair-site terms $Q_{XY}$ are rewritten as
\begin{eqnarray*}
Q_{XY} &=& \sum_{z=\pm,u} \big( P(X_1^+,Y_1^+,Z^z_1) 
- P(X_1^+, Y_2^+,Z^z_2) \\
&& 
- P(X_2^+,Y_1^+,Z^z_2)
- P(X_2^+,Y_2^+,Z^z_1)
\big).
\end{eqnarray*}
Here we take $Z_k$ for each $(X_i,Y_k)$ such that $k=1$ if $i=j=1$ or $i=j=2$, and $k=2$ if $i=1$ and $j=2$ or $i=2$ and $j=1$.
Then, the geometric tripartite extension of CH inequality is rewritten by
\begin{eqnarray*}
\label{eq:chmeb}
T_{111} + T_{122} + T_{212} + T_{221} \ge 0,
\end{eqnarray*}
where, for $P_{ijk}^{abc} =P(A_i^a,B_j^b,C_k^c)$,
\begin{eqnarray*}
\label{eq:chmeb111}
T_{111} &=& P_{111}^{+++} \nonumber \\
&& + P_{111}^{++-} + P_{111}^{+-+} + P_{111}^{-++} \nonumber \\
&& + P_{111}^{++u} + P_{111}^{+u+} + P_{111}^{u++} \nonumber \\
T_{122} &=& -P_{122}^{+-+} + P_{122}^{-+-} + P_{122}^{u+u} \nonumber \\
&& + P_{122}^{-+u} - P_{122}^{+u+} + P_{122}^{u+-} \nonumber \\
T_{212} &=& -P_{212}^{++-} + P_{212}^{--+} + P_{212}^{uu+} \nonumber \\
&& - P_{212}^{++u} + P_{212}^{-u+} + P_{212}^{u-+} \nonumber \\
T_{221} &=& -P_{221}^{-++} + P_{221}^{+--} + P_{221}^{+uu} \nonumber \\
&&  + P_{221}^{+-u} + P_{221}^{+u-} - P_{221}^{u++}.
\end{eqnarray*}
We replace $P_{ijk}^{abc}$ with the coincident counts $N_{ijk}^{abc}=N P_{ijk}^{abc}$ and obtain a 3-qubit Eberhard inequality.

\section{A method to obtain a new set of inequalities which can be reduced to Mermin inequalities for subsystems more than three}
\label{CH-Mermin}
In this subsection, we propose a method  to derive an inequality for four qubits and also show that the  extension to five qubits is also possible. Our derivation is based on the following identity,
\begin{eqnarray}
\label{idn}
P(X_{i1})+P(X_{i2})+P(X_{i3})=P(Y_{i4})+2P(Z_{i4}),
\end{eqnarray}
where $X_{ij}$ are events in the event space $F$ (see Sec.~\ref{sec:subII1}), $Y_{i4}=\oplus_{j=1}^3 X_{ij}$, and $Z_{i4}=(X_{i1}\cap X_{i2})\oplus (X_{i1}\cap X_{i3})\oplus (X_{i2}\cap X_{i3}).$  This applies to $i=1,2,3$. Thus we have a set of $3$ equations. In addition, we consider the $4$-th equation, 
\begin{eqnarray}
\label{idn1}
P(Y_{44})+2P(Z_{44})=P(Y_{41})+P(Y_{42})+P(Y_{43}),
\end{eqnarray}
where $Y_{4j}=\oplus_{i=1}^3 X_{ij}$,  $Y_{44}=\oplus_{j=1}^3Y_{4j},$ and $Z_{44}=(Y_{41}\cap Y_{42})\oplus (Y_{41}\cap Y_{43})\oplus (Y_{42}\cap Y_{43}).$ We sum the 4 equations in Eqs.~$\eqref{idn}$ and $\eqref{idn1}.$ Collecting the left hand sides together and similarly the right hand sides, we obtain a single equation. If $Z_{44} =\oplus_{i=1}^3 Z_{i4},$ then  
\begin{eqnarray}
\label{idn0}
\sum_{i=1}^3P(Z_{i4})\geq P(Z_{44}).
\end{eqnarray}
Hence, the summed equation can be written by an inequality,
\begin{eqnarray}
\label{idnn}
\sum_{i,j=1}^3 P(X_{ij}) + P(Y_{44}) \ge \sum_{k=1}^3 \left( P(Y_{k4}) + P(Y_{4k}) \right).
\end{eqnarray}

We apply the procedure from Eq.~\eqref{idn} to Eq.~\eqref{idnn} in order to derive multipartite  inequalities by replacing events $X_{ij}$ with symmetric differences between spatially separated subsystems. For example, let's consider $A_l,$ $B_m,$ $C_n,$  $D_p$ are events associated with the measurements of Alice, Bob, Charlie, and Daniel, respectively, where $l, m, n, p=1,2.$ Letting $\gamma_{lmnp}=A_l\oplus B_m\oplus C_n\oplus D_p$, the replacements 
 $$X_{11}\to \gamma_{1112}, ~X_{12}\to \gamma_{1121},~X_{13}\to \gamma_{1122},$$ $$X_{21}\to\gamma_{1221}, ~X_{22}\to\gamma_{1212},~ X_{23}\to \gamma_{1211},$$ $$X_{31}\to \gamma_{2121},~ X_{32}\to \gamma_{2112},~ X_{33} \to \gamma_{2211},$$ in Eq.~\eqref{idnn}  yield an inequality for four qubits, as they satisfy Eq.~\eqref{idn0}.

5-qubit inequality is derived, as done for four qubits. We define $\gamma_{lmnpr}=A_l\oplus B_m\oplus C_n\oplus D_p\oplus E_r,$ where the extra events $E_r$ belong to Eve for $r=1, 2.$ The inequality for  five qubits results from inequality~\eqref{idnn} by replacing
$$X_{11}\to \gamma_{11122}, ~X_{12}\to \gamma_{11212},~X_{13}\to \gamma_{11221},$$ $$X_{21}\to\gamma_{12211}, ~X_{22}\to\gamma_{12121},~ X_{23}\to \gamma_{12112},$$ $$X_{31}\to \gamma_{21211},~ X_{32}\to \gamma_{21121},~ X_{33} \to \gamma_{22111}.$$
Here, we have used $Z_{44} = \oplus_{i=1}^3 Z_{i4}$ and Eq.~\eqref{idn0}. In addition one can apply the procedure in Sec.~\ref{rm} to derive Mermin inequalities for four and five qubits by using inequality~\eqref{idnn}.

We remark that the inequality presented in the section is derived by a vector of events $X_i$ for two and three qubits, and the one by a matrix of events $X_{ij}$ for four [Eq.~$\eqref{idnn}$] and five qubits. It is conjectured that one needs an $n$-th rank tensor of events $X_{i_1 i_2 \cdots i_n}$ for the specific type of inequalities of $2n$ and $2n+1$ qubits. This is beyond the scope of this work.

\end{document}